\documentclass[11pt]{article}

\usepackage{hyperref}
\usepackage{amsmath,amsfonts,amssymb}
\hypersetup{dvips,dvipdfm,colorlinks=true,urlcolor=magenta,filecolor=magenta,linktoc=page,citecolor=red,linkcolor=blue,bookmarks=true}
\usepackage{graphicx,epstopdf}
\usepackage{multirow,array}
\begin{document}
	\title{\textbf{A study of Raychaudhuri equation and geodesic focusing in Fractal Universe}}
	\author{Madhukrishna Chakraborty \footnote{\url{ chakmadhu1997@gmail.com} (corresponding author)}~~ and~~
		Subenoy Chakraborty\footnote{\url{schakraborty.math@gmail.com}}
		\\ \textit{\scriptsize Department of Mathematics, School of Basic and Applied Sciences, Adamas University, Kolkata-700126, West Bengal, India}}
	\date{}
	\maketitle
	\begin{abstract}
		The paper deals with the modified Raychaudhuri equation (RE) within the framework of homogeneous and isotropic Fractal Universe. Focusing of a congruence of time-like geodesics has been examined for three generic choices of the fractal function. Finally, comments on the existence and possible avoidance of the initial big-bang singularity have been made by examining the sign of convergence scalar in the fractal models under consideration.
		\end{abstract}
	\section{Introduction}
	Till date, General Relativity (GR) \cite{Wald:1984rg}, \cite{Weinberg:1972kfs} is the most well-accepted theory of gravity that successfully describes the universal dynamics. The popularity enhanced after the detection of gravitational waves \cite{LIGOScientific:2017vwq}. However, there are a few drawbacks of GR. Existence of space-time singularities is the most crucial one among them. The inevitable existence of singularity in GR has been proved in the seminal singularity theorems by Penrose and Hawking \cite{Hawking:1973uf}, \cite{Penrose:1964wq}, \cite{Hawking:1970zqf} for which in 2020, Penrose has been awarded the Noble Prize. The key argument presented in the proof is focusing of a congruence of geodesics and Raychaudhuri equation \cite{akr}. According to the Focusing theorem, an initially converging congruence of null (time-like) geodesics will focus within a finite value of affine parameter (proper time) provided the matter content of the universe is usual or normal satisfying strong energy condition (SEC). FT is the most crucial consequence of RE \cite{Kar:2006ms}, \cite{Chakraborty:2024khs}. RE is a first order non-linear differential equation in the expansion scalar corresponding to a bundle of geodesics (time-like/null). In other words, it is the evolution equation for geodesics. This is the reason why RE plays a vital role in cosmic evolution \cite{Chakraborty:2024wty}. RE is widely used to find exact solutions in GR. Hence it finds application in cosmology and astrophysics. Analysis of singularity in cosmological model via geodesic focusing is an important application of RE \cite{Chakraborty:2024obs}. There are other well-known and in some sense related approaches to resolution of initial time singularity. For ref. see \cite{Gorji:2014pka}. Further, RE plays a very crucial role in the formation and stability analysis of compact stellar objects. For realistic anisotropic charged model, it is necessary to check the equilibrium, stability, causality, regularity and energy conditions of stellar objects. The present study may be applied the evolution of the compact object under various backgrounds \cite{Khan:2024jnb}. In this context, RE in anisotropic background might be helpful \cite{Chakraborty:2023rgb}.  Stability of compact stellar structures and complex cosmological structures in modified gravity theories can be ensured by the RE \cite{Yousaf:2021xex}, \cite{Yousaf:2021rws}, \cite{Yousaf:2019pqe}, \cite{Khan:2024vsh}. RE in static spherically symmetric background has been studied in \cite{Chakraborty:2024dpx}. This study can be extended in case of  charged cylindrical gravastar-like structures \cite{Yousaf:2020srr}.  Formation of Neutron star (NS) under the effect of RE coupled with Tolman–Oppenheimer–Volkoff equation (TOV) equation has been attempted in \cite{Debnath:2021kyq}. Along with the TOV equation, RE may provide structural stability of the surrounding space-time from further collapse to singularity for a newly born NS. Evolution equation for static stellar model can be constructed out of the RE which provides sufficient stability to maintain hydro-static equilibrium of the star against gravitational collapse. RE can be used to consider the stability condition for a newly born stellar structure in its initial evolutionary stage and also in the final state of its collapse \cite{Choudhury:2019snb}, \cite{Kouretsis:2010nu}. Geodesic congruences and collapsing stellar distribution has been studied in $f(T)$ theories \cite{Chakrabarti:2020ngy}. RE may be applied to stability analysis of the compact star configurations indicating the potential applicability of different background geometries in understanding and characterizing celestial bodies in astrophysics \cite{Albalahi:2024ujg}, \cite{Yousaf:2024src}, \cite{Albalahi:2024vpy}. Physical features like regularity at the center, the measure of anisotropy, energy conditions, stability, causality and well behaved condition of compact stars can be analyzed from the point of view of RE. Energy conditions play a very crucial role in understanding the causal and geodesic behavior of space-time. These conditions are derived from the well known RE. The modeling of self-gravitating compact configurations using radial metric deformation approach has been performed in \cite{Yousaf:2024fkr}. RE may be applied for the stability analysis in the constructed model.
	
	 It is interesting to note that RE is purely a geometric identity in the Lorentzian geometry and has nothing to do with any gravity theory. Thus, the scope of RE is not only restricted to gravity and cosmology but beyond that. Nevertheless, RE has a lot of applications in gravity, cosmology and astrophysics.  RE is used to study different phenomena like cosmological evolution \cite{Chakraborty:2024wty}, focusing of congruence \cite{Choudhury:2021huy}, \cite{Chakraborty:2023voy},  gravitational collapse \cite{Choudhury:2019snb} etc. To find its importance in gravity and cosmology one has to invoke the Einstein's field equations which associate the geometry of space-time with the matter content of the space-time. RE guarantees the fact that focusing is inevitable in GR. However, focusing alone does not lead to space-time singularity. Focusing forms a caustic which is a point at which all geodesics focus. A caustic may be regarded as a singularity of the congruence. Thus, a congruence singularity may not be a singularity of the space-time. The notion of focusing together with a few reasonable conditions on space-time enforce the general existence of curvature singularities. In GR, SEC applied to Einstein's field equations leads to Convergence Condition (CC). This hints the attractive nature of gravity and consequently leads to geodesic focusing. In other extended theories of gravity, the field equations are different from that in GR. Thus, there is a possibility for violation of CC even with the assumption of SEC on matter. This might lead to possible avoidance of singularity in some cases. The main goal of the study is to analyze the behavior of geodesics in fractal universe using a modified gravitational model, called fractal gravity via the modified RE in fractal grvaity. In this context, readers may refer to RE in modified gravity theories like $f(R)$ gravity \cite{Choudhury:2019zod}, \cite{Chakraborty:2023ork}; $f(T)$ gravity \cite{Chakraborty:2023yyz}; scalar tensor theories \cite{Choudhury:2021zij}; string theory \cite{Burger:2018hpz}; braneworld gravity \cite{Ghosh:2010gq},; $f(G,T^{2})$ gravity \cite{Yousaf:2021xex}; $f({\mathbb {G}}, T)$ gravity \cite{Bhatti:2021pxr} and loop quantum cosmology \cite{Blanchette:2020kkk}
	
	In the present work, we study the focusing of geodesics in Fractal Universe \cite{Asghari:2022kia},  \cite{Bose:2022wla} via the celebrated RE. The motivation behind this is to find out whether CC is universally satisfied like GR or not and to examine the fractal effects on focusing of a congruence. CC has been analyzed for three generic choices of the fractal function which are cosmologically viable and consistent with latest observations \cite{El-Nabulsi:2024zqt}, \cite{Das:2018bxc}, \cite{Pawar:2024juv}. In this regard, possible existence and avoidance of the initial big-bang singularity have been investigated by examining the signature of the convergence scalar evaluated for the fractal models. Actually, fractal universe can be considered as standard cosmology with two interacting fluids \cite{Das:2018bxc}. In \cite{Das:2018bxc}, fractal cosmology has been shown to be same as a particle creation mechanism in the context of non-equilibrium thermodynamic prescription. Observational constraints on fractal gravity has been performed in \cite{Pawar:2024juv}. Warm inflationary scenario in fractal universe has been studied in \cite{Bose:2022wla}. For constant fractal function, the standard Friedmann equations are recovered. Thus, a fractal gravity theory can be considered as an extension of GR. It is interesting to note that Fractal structure has connection with Dispersion Relations and therefore fundamental constant of nature. In this context the authors in \cite{Mehdipour:2008kno} considered the problem of wave packet broadening in the framework of the
		Generalized Uncertainty Principle (GUP) of quantum gravity. They have found a fractal Klein Gordon equation to further analyze the wave packet broadening in a foamy (fuzzy) spacetime and
		derived a Modified Dispersion Relation in the context of GUP which shows an extra
		broadening due to gravitational induced uncertainty. As a result of these dispersion relations, a
		generalized Klein-Gordon equation has been obtained by them \cite{Mehdipour:2008kno}.

	 The layout of the paper is as follows: Section 2 deals with an overview of RE and FT in GR. Section 3 deals with a brief review of fractal gravity. RE and modified CC in Fractal universe have been presented in Section 4 subject to three choices of the fractal function in the corresponding subsections. Finally, the paper ends with concluding remarks in Section 5.
	 \section{An overview of RE and FT in GR}
	 RE is a very well-known and useful equation in GR and cosmology. Actually, RE describes the kinematics of a deformable medium containing a bundle of time-like/ null geodesic. It is a first order non linear differential equation in the expansion scalar representing the congruence. Consider a congruence of time-like geodesic having velocity vector $u^{a}$, then RE is given by \cite{Chakraborty:2023rgb}, \cite{Chakraborty:2023lav}
	 \begin{equation}
	 	\dfrac{d\Theta}{d\tau}=-\dfrac{\Theta^{2}}{3}-2\sigma^{2}+2\Omega^{2}-R_{ab}u^{a}u^{b}\label{eq1}
	 \end{equation} where $\Theta=\nabla_{a}u^{a}$ is the expansion scalar that measures the relative expansion or contraction of the bundle of geodesics, $\sigma_{ab}=\nabla_{(_{b}u_{a})}-\dfrac{1}{3}h_{ab}\Theta=\nabla_{b}u_{a}+\nabla_{a}u_{b}-\dfrac{1}{3}h_{ab}\Theta$ is the $(0,2)$ shear tensor, $2\sigma^{2}=\sigma_{ab}\sigma^{ab}$ is the anisotropy scalar that measures the kinematic ansiotropy, $h_{ab}=g_{ab}+u_{a}u_{b}$ is the induced spatial metric, $\Omega_{ab}=\nabla_{[_{b}u_{a}]}=\nabla_{b} u_{a}-\nabla_{a} u_{b}$ is the $(0,2)$ vorticity tensor, $2\Omega^{2}=\Omega_{ab}\Omega^{ab}$ is the vorticity scalar that measures the rotational behavior of the congruence, $\tau$ is the proper time, $R_{ab}$ is the Ricci tensor projected along the geodesics. The term $R_{ab}u^{a}u^{b}$ encapsulates the contribution of space-time geometry and plays a significant role in focusing.
 
 For hyper-surface orthogonal congruence of time-like geodesics, the vorticity vanishes by virtue of Frobenius theorem of differential geometry and the equation (\ref{eq1}) reduces to much simplified form as
 \begin{equation}
 		\dfrac{d\Theta}{d\tau}=-\dfrac{\Theta^{2}}{3}-2\sigma^{2}-R_{ab}u^{a}u^{b}\label{eq2}
 \end{equation}
If we consider GR then SEC on matter implies $T_{ab}u^{a}u^{b}+\dfrac{1}{2}T\geq0$ or $R_{ab}u^{a}u^{b}\geq0$ by virtue of Einstein's field equations. Consequently, the r.h.s of the above equation (\ref{eq2}) is negative as $\sigma_{ab}\sigma^{ab}\geq0$ for spatial $\sigma_{ab}$. This shows that if we consider an initially converging congruence i.e, $\Theta_{0}=\Theta(0)<0$, then geodesics will focus within a finite value of the proper time $\tau\leq 3 \Theta_{0}^{-1}$. This concept is popularly known as FT and mathematically this focusing condition can be written as \cite{Chakraborty:2024obs}
\begin{equation}
	\dfrac{d\Theta}{d\tau}+\dfrac{\Theta^{2}}{3}\leq0\label{eq3}
\end{equation} i.e, $R_{ab}u^{a}u^{b}\geq0$. This is known as CC. It is to be noted that focusing alone does not always lead to space-time singularity. A few physically reasonable conditions need to be satisfied in order that this congruence singularity can be a space-time or Black-hole singularity. On the other hand, violation of the focusing condition comes up with the possible avoidance of singularity. Precisely speaking, if SEC is violated CC no more holds and focusing can be prevented. This is because, an important assumption behind the FT is $R_{ab}u^{a}u^{b}\geq0$ and again this comes from the SEC. Therefore, when the additional term in the energy-momentum tensor for modified gravity theory is considered then there is a possibility of avoidance of singularity. 

In FLRW background, $\sigma=0$, $\Theta=3H=3\dfrac{\dot{a}}{a}$ and $R_{ab}u^{a}u^{b}=\dfrac{(\rho+3p)}{2}$. Thus, the RE in FLRW background can be written as 
\begin{equation}
	\dfrac{\ddot{a}}{a}=-\dfrac{(\rho+3p)}{6}
\end{equation}
If $(\rho+3p)\geq0$, then $\ddot{a}\leq0$ and vice-versa. Thus, in order to accommodate the present accelerated expansion of the universe into the theoretical framework, notion of Dark energy comes up satisfying $(\rho+3p)\leq0$. This shows that, RE is essential in hinting the exotic matter or dark energy which is believed to be responsible for the late time acceleration.
\section{Fractal gravity: Basic equations}
 With the assumption that matter is minimally coupled to gravity, the sum of action of Einstein gravity in a fractal space-time is given by  \cite{Bose:2022wla}, \cite{El-Nabulsi:2024zqt}, \cite{Das:2018bxc}.
 \begin{equation}
 	\mathcal{S}=\mathcal{S_{M}}+\mathcal{S}_{g}\label{eq3}
 \end{equation}
where $\mathcal{S_{M}}$ is the matter part of the action and is given by
\begin{equation}
	\mathcal{S_{M}}=\int d^{4}x~v(x)~\sqrt{-g}~\mathcal{L_{M}}
\end{equation} and $\mathcal{S}_{g}$ is the gravitational part of the action given by 
\begin{equation}
	\mathcal{S}_{g}=\dfrac{1}{16\pi G}\int d^{4}(x)~v(x)~\sqrt{-g}~\left(R-w~\partial_{a}v~\partial^{a}v\right)\label{eq6}
\end{equation} where $g$ is the determinant of $g_{ab}$, $R$ is the Ricci scalar, $v$ is the fractal function, $w$ is the fractal parameter, $d^{4}x$ is the standard measure. It is replaced by a Lebesgue-Stieltjes measure $d\bf{g}(x)$ such that the scaling dimension of $\bf{g}$ is $-4\alpha$, ($0<\alpha<1$). $\alpha$ is the parameter that denotes the fraction of states preserved at a given time when a system evolves. Further, it is to be noted that the measure $v$ is neither a scalar field nor any dynamical object. Its profile is fixed by the underlying geometry on a prior basis. Varying the action w.r.t homogeneous, isotropic and flat FLRW metric $g_{ab}$, the field equations (modified Friedmann equations) in a fractal universe can be obtained as (choosing $8\pi G=1$) \cite{Bose:2022wla}, \cite{El-Nabulsi:2024zqt}, \cite{Das:2018bxc}.
\begin{eqnarray}
	3H^{2}=\rho-3H\dfrac{\dot{v}}{v}+\dfrac{w}{2}\dot{v}^{2}\label{eq8}\\
	2\dot{H}+3H^{2}=-p-2H\dfrac{\dot{v}}{v}-\dfrac{w}{2}\dot{v}^{2}-\dfrac{\ddot{v}}{v}\label{eq9}
\end{eqnarray} where $a(t)$ is the cosmic scale factor, $H=\dfrac{\dot{a}}{a}$ is the Hubble parameter, $\rho$ is the energy density, $p$ is the pressure of the barotropic fluid component and `~.~' represents differentiation w.r.t cosmic time $t$. Clearly, $v=$ constant yields the standard Friedmann equations. The continuity equation in fractal universe takes up the form \cite{Bose:2022wla}, \cite{El-Nabulsi:2024zqt}, \cite{Das:2018bxc}.
\begin{equation}
	\dot{\rho}+\left(3H+\dfrac{\dot{v}}{v}\right)(p+\rho)=0\label{eq10}
\end{equation} and the expression for $\rho$ is given by
\begin{equation}
	\rho=\rho_{0}(a^{3}v)^{-(1+\omega)}
\end{equation} where a barotropic equation of state i.e, $p=\omega \rho$ is assumed ($\omega$, a constant).
\section{RE in Fractal universe: Analysis of CC}
\subsection{General Formulation}
From the Friedmann equations (\ref{eq8}) and (\ref{eq9}), one obtains the modified RE in Fractal universe as
\begin{equation}
	\dfrac{\ddot{a}}{a}=-R_{1}+R_{2}
\end{equation}
 where
 \begin{eqnarray}
 	R_{1}=\dfrac{(\rho+3p)}{6}=\dfrac{(1+3\omega)}{6}\rho_{0}(a^{3}v)^{-(1+\omega)}\\
 	R_{2}=-\dfrac{\dot{a}\dot{v}}{2av}-\dfrac{1}{2}\dfrac{\ddot{v}}{v}-\dfrac{1}{3}\dot{v}^{2}
 \end{eqnarray}
In order to avoid focusing, we need to have $-R_{1}+R_{2}>0$ i.e, $R_{2}>R_{1}$. If matter satisfies SEC, then $R_{1}\geq0$ for all $t$. Thus, $R_{2}\geq0$ and hence for expanding model of universe ($\dot{a}>0$) one should have $\dot{v}\leq0$ i.e, the fractal function must be a decreasing function of the cosmic time $t$.  Thus, we have to examine the sign of $R_{2}$ and its role in the overall sign of $-R_{1}+R_{2}$ to study the focusing of a congruence of time-like geodesic in Fractal universe. The effect of fractal gravity is present in both the terms $R_{1}$ and $R_{2}$ via the fractal function $v$. Thus, we study the CC in the following fractal models explicitly subject to some generic choices of the fractal function given in literature. Let us denote $-R_{1}+R_{2}$ by $\mathcal{R}$ and call it a convergence scalar. It may be noted that, CC holds for $\mathcal{R}\leq0$ and focusing is prevented if $\mathcal{R}\geq0$.
\subsection{Choice 1: $v=v_{0}t^{-\beta},~~\beta=4(1-\alpha)$ ; $a=a_{0}t^{m}$}
\begin{figure}[h!]
	\begin{minipage}{0.3\textwidth}
		\centering\includegraphics[height=5cm,width=5cm]{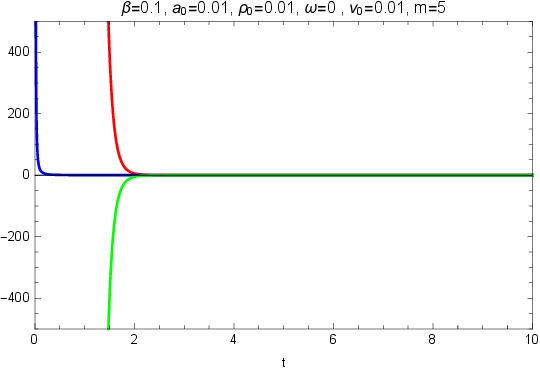}
	\end{minipage}~~~
	\begin{minipage}{0.3\textwidth}
		\centering\includegraphics[height=5cm,width=5cm]{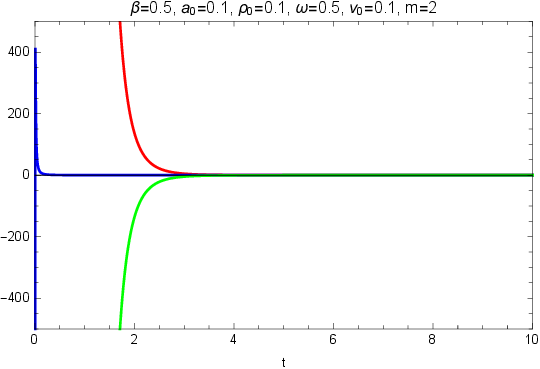}
	\end{minipage}\hfill
	\begin{minipage}{0.3\textwidth}
		\centering\includegraphics[height=5cm,width=5cm]{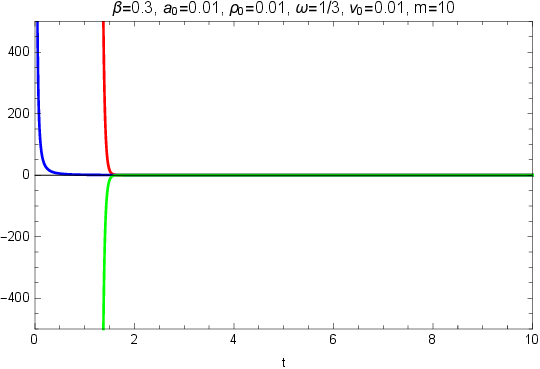}
	\end{minipage}\hfill
	\begin{minipage}{0.3\textwidth}
		\centering\includegraphics[height=5cm,width=5cm]{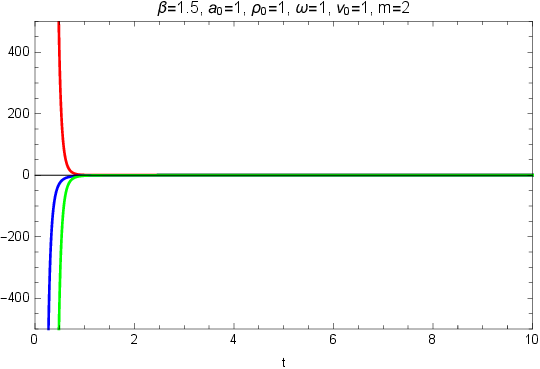}
	\end{minipage}~~~~
\begin{minipage}{0.3\textwidth}
	\centering\includegraphics[height=5cm,width=5cm]{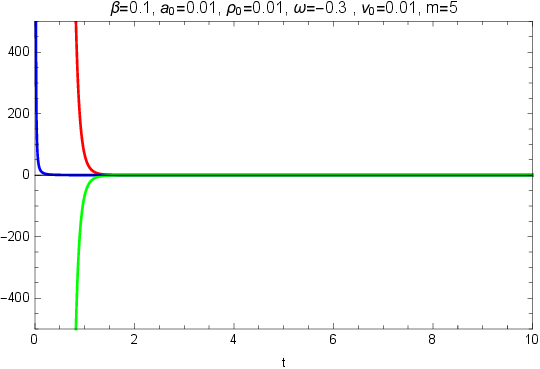}
\end{minipage}~~~~~~~~~~
	\begin{minipage}{0.3\textwidth}
	\centering\includegraphics[height=5cm,width=5cm]{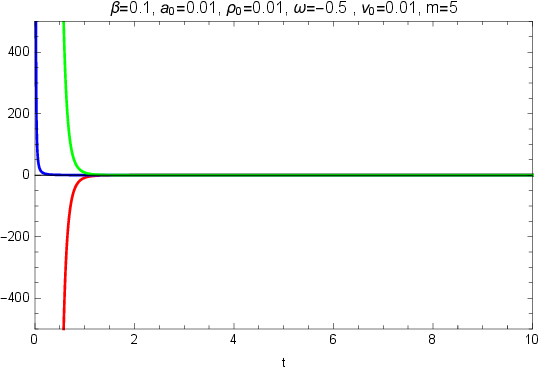}
\end{minipage}
\caption{Time variation of $R_{1}$ (red), $R_{2}$ (blue) and $\mathcal{R}$ (green) for choice 1 with various choices of the parameters specified in each panel}\label{f1}
\end{figure}

This choice is one of the most common and popular choices of the fractal function found in literature  \cite{Asghari:2022kia},  \cite{Bose:2022wla}. In this choice, $\alpha>0$ is associated with the Hausdorff dimension of the physical space-time.  It neither recovers the standard measure of GR nor does it represent any multifractal geometry. For this choice, the expressions for $R_{1}$ and $R_{2}$ are given by 
\begin{eqnarray}
	R_{1}=\dfrac{(1+3\omega)}{6}~\rho_{0}~(a_{0}^{3}v_{0})^{-(1+\omega)}~t^{(\beta-3m)(1+\omega)}\\
	R_{2}=\dfrac{\beta(m-\beta-1)}{2t^{2}}-\dfrac{\beta^{2}v_{0}^{2}}{3t^{2(\beta+1)}}
\end{eqnarray}
The time variation of $R_{1},~R_{2}$ and $\mathcal{R}$ are shown in FIG. (\ref{f1}) for Choice 1. In all plots, $R_{1}\geq0$ (red) except the last one. For higher positive values of $\beta=4(1-\alpha)$, $R_{2}\leq0$.
\subsection{Choice 2: $v=v_{0}a^{m}~,m<0$}
\begin{figure}[h!]
	\begin{minipage}{0.3\textwidth}
		\centering\includegraphics[height=5cm,width=5cm]{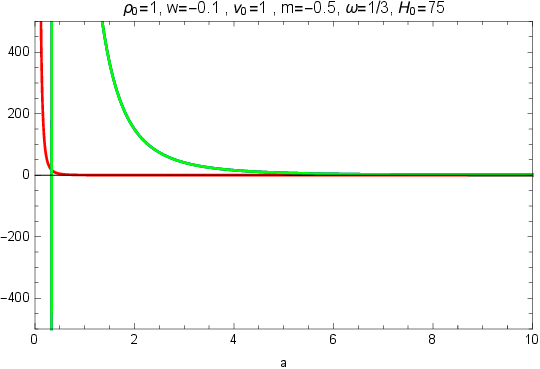}
	\end{minipage}~~~~
	\begin{minipage}{0.3\textwidth}
		\centering\includegraphics[height=5cm,width=5cm]{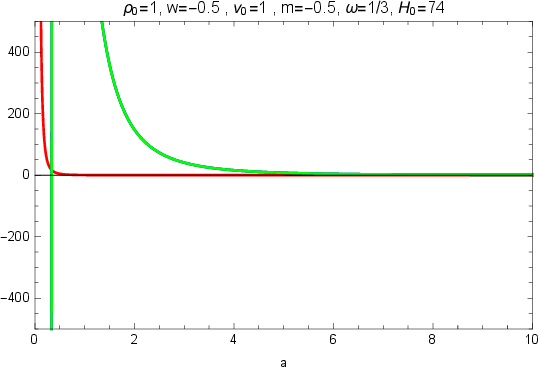}
	\end{minipage}~~~~~~~
	\begin{minipage}{0.3\textwidth}
		\centering\includegraphics[height=5cm,width=5cm]{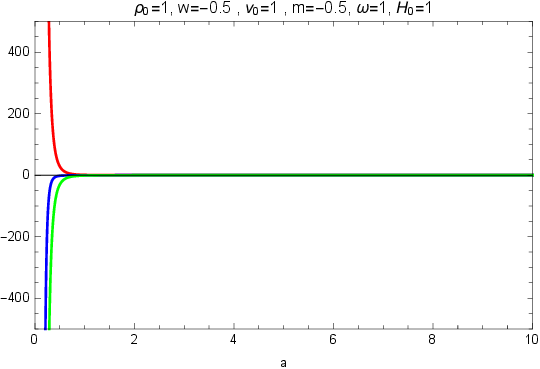}
	\end{minipage}
	\caption{Variation of $R_{1}$ (red), $R_{2}$ (blue) and $\mathcal{R}$ (green) w.r.t cosmic scale factor $a$ for choice 2 with various choices of the parameters specified in each panel}\label{f2}
\end{figure}

This choice is a monomial of the scale factor and reduces to Choice 1 in the matter dominated era as in that case the scale factor grows in power-law form \cite{El-Nabulsi:2024zqt}, \cite{Das:2018bxc}. Hence, the parameter $m$ is related to the  Hausdorff dimension of the physical space-time. This choice, if put in the energy conservation equation (\ref{eq10}) yields
\begin{equation}
	\rho=\rho_{0}a^{-(1+\omega)(m+3)}
	\end{equation}
and consequently from the field equation one gets
\begin{equation}
	H=H_{0}\sqrt{\dfrac{a^{-(1+\omega)(m+3)}(6(1+m)-wm^{2}v_{0}^{2})}{(6(1+m)-wm^{2}v_{0}^{2}a^{2m})}}
\end{equation}
where $H_{0}$ is the present value of the Hubble parameter. The explicit expressions for $R_{1}$ and $R_{2}$ are given by
\begin{eqnarray}
R_{1}=	\dfrac{(1+3\omega)}{3(2+m)}\rho_{0}~a^{-(1+\omega)(m+3)}\\
R_{2}=-mH^{2}\left(\dfrac{1}{2}+\dfrac{1}{3}mv_{0}^{2}a^{2m}\right)+\dfrac{1}{2}\dfrac{m(1-m)H}{a}
\end{eqnarray}
The variation of $R_{1},~R_{2}$ and $\mathcal{R}$ w.r.t scale factor are shown in FIG. (\ref{f2}) for Choice 2. For this choice of the fractal function, CC is not universally satisfied (as sign of $\mathcal{R}$ is indefinite) if we consider present value of the Hubble parameter. However, if the value of Hubble parameter is very small, then focusing can not be prevented. This cosmological model is consistent with latest observations \cite{El-Nabulsi:2024zqt}, \cite{Das:2018bxc}, \cite{Pawar:2024juv}
\subsection{Choice 3: $v=v_{0}\exp(-\beta t)$, $a=a_{0}\exp(\mu t)$}
\begin{figure}[h!]
 	\begin{minipage}{0.3\textwidth}
 	\centering\includegraphics[height=5cm,width=6cm]{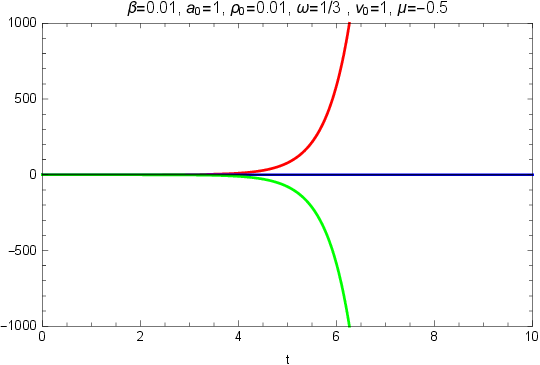}
 \end{minipage}\hfill
 \begin{minipage}{0.3\textwidth}
 	\centering\includegraphics[height=5cm,width=6cm]{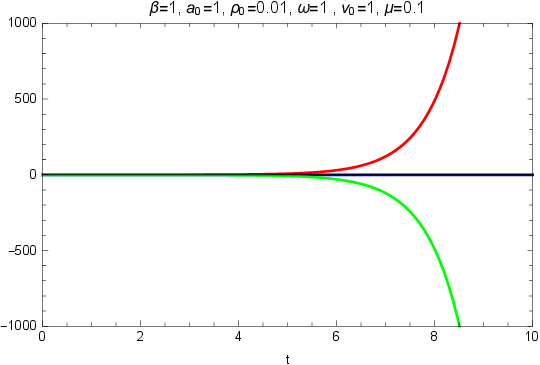}
 \end{minipage}
 \caption{ Time Variation of $R_{1}$ (red), $R_{2}$ (blue) and $\mathcal{R}$ (green) for choice 3 with various choices of the parameters specified in each panel}\label{f3}
\end{figure}

This choice is an exponential form of the fractal function. Since Choice 1 neither represents multifractal geometry nor recovers GR so in literature this exponential choice of fractal function is used \cite{Bose:2022wla}. For this choice, $R_{1}$ and $R_{2}$ assume the following forms
\begin{eqnarray}
	R_{1}=\dfrac{(1+3\omega)}{6}~\rho_{0}~(a_{0}^{3}v_{0})^{-(1+\omega)}~\exp{(\beta-3\mu)(1+\omega)t}\\
	R_{2}=\dfrac{1}{2}\beta(\mu-\beta)-\dfrac{1}{3}~v_{0}^{2}~\beta^{2}\exp(-2\beta t)
\end{eqnarray}
 Time variation of $R_{1},~R_{2}$ and $\mathcal{R}$ are shown in FIG. (\ref{f3}) for Choice 3. In this case, effect of $R_{2}$ is very negligible and thus $R_{1}$ and $\mathcal{R}$ are mirror images of each other. This clearly, shows that exotic matter can prevent focusing and it is not possible to avoid singularity with usual matter in this particular fractal model irrespective of the choices of $\beta$ and $\mu$.
\section{Conclusion}
The paper shows how the RE is modified in Fractal universe and how the geodesics behave in the framework of homogeneous and isotropic Fractal gravity theory. RE together with FT marks the inevitable existence of singularity in GR. In the present work, we have attempted to answer the question ``Is the initial singularity inevitable in fractal gravity theory?". In literature, RE in modified theories of gravity show that singularity is not inevitable but may be avoided under some physically reasonable criteria. On the other hand, fractal gravity theory has attracted interest in literature in the recent past. This motivates us to formulate the modified RE and the corresponding CC in fractal gravity. This also helps us to understand the notion of geodesic focusing or precisely, the behavior of geodesics in fractal universe. For this, we have considered three choices of the fractal function and examined the sign of the convergence scalar $\mathcal{R}$. Clearly, the three generic choices of the fractal function imply that $v$, the fractal function is some power of $a$, the cosmic scale factor. In case of choices 1 and 3 $a$ is specified while choice 2 is the most general one. It is to be noted that Choice 2 i.e, the most general choice is cosmologically viable and supports the latest observational data \cite{El-Nabulsi:2024zqt}, \cite{Das:2018bxc}, \cite{Pawar:2024juv}. This is the motivation to work with these viable models. Although the present work is theoretical but the models under consideration are consistent with observation. The signature of the Raychaudhuri scalar determines whether a congruence of geodesic will focus or not. If the signature of convergence scalar is positive definite then focusing will occur. This is called the Focusing theorem which follows as a consequence of the Raychaudhuri equation.  Focusing or convergence may or may not lead to formation of singularity. However, if one can avoid focusing then there will be no singularity. This concept is used in the present fractal models. In these models we have evaluated the convergence scalar and studied its signature. This helped us to comment on the existence and possible avoidance of singularity in the fractal models.

For Choice 1 i.e, a power law choice of the fractal function in cosmic time $t$ it is found that, if matter satisfies SEC ($R_{1}\geq0$), then focusing can not be prevented as $\mathcal{R}\leq0$ from FIG. (\ref{f1}). The power law form of scale factor shows that the model has an initial singularity. Thus, in this model focusing occurs and there is a singularity of the space-time. However, exotic matter (matter violating SEC or $R_{1}\leq0$) can avoid this singularity.

For Choice 2, a monomial in scale factor $a$ has been considered as the fractal function. This choice, gives $\mathcal{R}$ in terms of $a$.  FIG. (\ref{f2}) shows that CC is not universally satisfied in this fractal model as signature of $\mathcal{R}$ is indefinite if we consider the present value of the Hubble parameter. However, for small value of the Hubble parameter it is found that CC holds and singularity is inevitable even with usual matter. Thus, in this model singularity may be avoided with suitable choices of the parameters involved.

For Choice 3 which is an exponential choice of the fractal function it is seen from FIG. (\ref{f3}) that focusing can not be prevented with usual matter. The corresponding choice of the scale factor shows that, there is no singularity of the space-time but still focusing occurs. This proves the fact that even if focusing occurs the space-time may not develop a singularity.

Thus in fractal universe, focusing may occur with or without a singularity. However, since CC is not universally satisfied for the second choice of the fractal function with normal matter, therefore there is a chance of possible avoidance of singularity in this fractal model.
\section*{Acknowledgment}
 The authors thank the anonymous reviewers for their valuable and insightful comments that improved the quality and visibility of the work. The authors also thank Department of Mathematics, School of Basic and Applied Sciences, Adamas University for providing research facilities.

	\end{document}